# Computing Grain Boundary 'Phase' Diagrams

Jian Luo[*]

Department of Nanoengineering; Program of Materials Science and Engineering, University of California San Diego, La Jolla, California, U.S.A.

**Abstract**

Grain boundaries (GBs) can be treated as two-dimensional (2-D) interfacial phases (also called 'complexions') that can undergo interfacial phase-like transitions. As bulk phase diagrams and calculation of phase diagram (CALPHAD) methods are a foundation for modern materials science, we propose to extend them to GBs to have equally significant impacts. This perspective article reviews a series of studies to compute the GB counterparts to bulk phase diagrams. First, a phenomenological interfacial thermodynamic model was developed to construct GB lambda diagrams to forecast high-temperature GB disordering and related trends in sintering and other properties for both metallic and ceramic materials. In parallel, an Ising-type lattice statistical thermodynamic model was utilized to construct GB adsorption diagrams, which predicted first-order GB adsorption (*a.k.a.* segregation) transitions and critical phenomena. These two simplified thermodynamic models emphasize on the GB structural (disordering) and chemical (adsorption) aspects, respectively. Subsequently, hybrid Monte Carlo and molecular dynamics atomistic simulations were used to compute more rigorous and accurate GB 'phase' diagrams. Computed GB diagrams of thermodynamic and structural properties were further extended to include mechanical properties. Moreover, machine learning was combined with atomistic simulations to predict GB properties as functions of four independent compositional variables and temperature in a 5-D space for a given GB in high-entropy alloys or as functions of five GB macroscopic (crystallographic) degrees of freedom (DOFs) plus temperature and composition for a binary alloy in a 7-D space. Relevant other studies are examined. Future perspective and outlook, including two emerging fields of high-entropy grain boundaries (HEGBs) and electrically (or electrochemically) induced GB transitions, are discussed.

---

[*] Email: jluo@alum.mit.edu

# Graphical Abstract

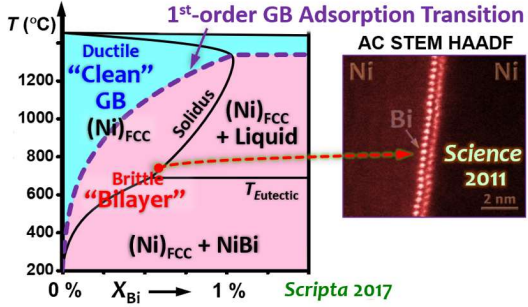
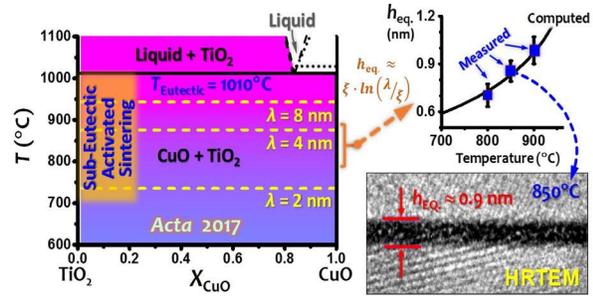
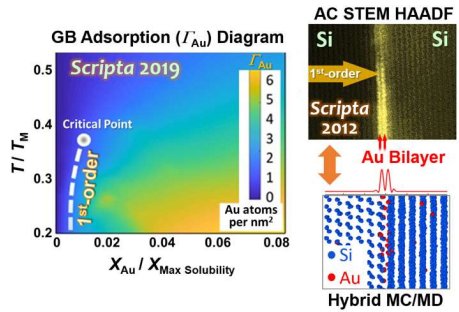
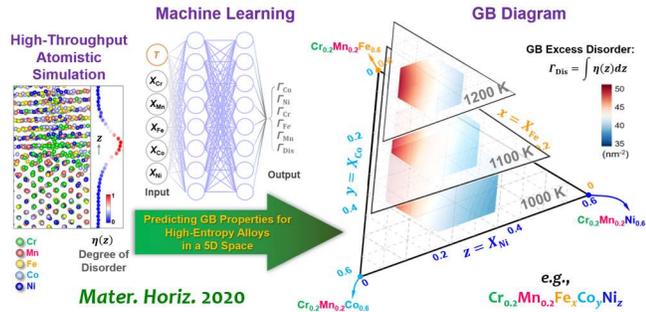

As bulk phase diagrams are a foundation for materials science, computing their grain boundary counterparts that can have equally significant impacts.

# 1. Introduction

Phase transition (or transformation) is one of the most interesting physical phenomena, and it is of critical importance for materials science. Ice melts at 0 °C at 1 atm, which represents one well-known phase transition. On September 8, 1842, Michael Faraday noted in his dairy that the surface of ice can start to "melt" below 0 °C (Figure 1),[1,2] a phenomenon called "surface melting" or "premelting".[2-4] Furthermore, Faraday used premelting to explain the fact that two blocks of ice can freeze together and a snowball can consolidate below 0 °C,[2-4] which are examples of sintering. Interestingly, our studies attributed the origin of solid-state activated sintering in ceramics[5,6] and refractory metals[7-11] to the enhanced mass transport in premelting like interfacial phases that are stabilized below the bulk solidus temperatures (Figure 1), which shed light on a long-standing mystery in materials science.

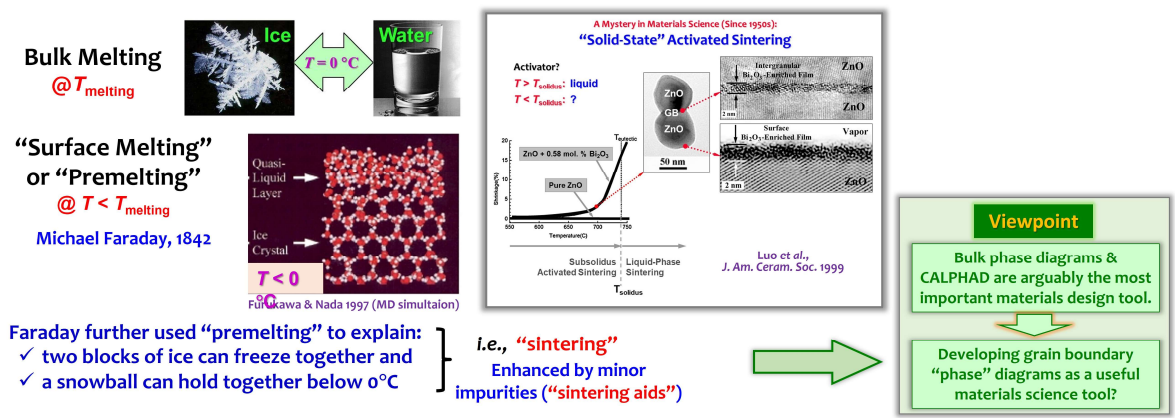

**Figure 1.** While ice melts at 0°C, Michael Faraday recognized that a surface layer of the ice can melt at tens of degrees below zero, a phenomenon called "surface melting" or "premelting".[1,2] A series of studies[6-12] attributed the origin of solid-state activated sintering in ceramics and refractory metals to enhanced mass transport in premelting like interfacial phases that are stabilized below the bulk solidus temperatures. Since bulk phase diagrams are one of the most useful tools for materials science, we envisioned that development of their GB counterparts can be equally important.

Materials scientists have long recognized that grain boundaries (GBs) can be treated as interfacial phases that are thermodynamically two-dimensional (2-D).[13,14] In 1968, Hart first proposed to treat GBs as 2-D interfacial phases.[15,16] Subsequent models developed by Hondros and Seah,[17,18] Cahn,[19-22] Clarke,[23,24] Carter *et al.*,[25-30] Wynblatt and Chatain,[31-33] Mishin *et al.*,[34-38] and Luo *et al.*[8,13,39,40] further elaborated the relevant concepts and phenomena. Notably, ceramic researchers have observed the widespread existence of a unique class of impurity-based intergranular (glassy) films (IGFs).[13,23,41,42] These IGFs can be equivalently understood to be (1) liquid-like interfacial films that adopt a nanoscale *equilibrium* thickness (the Clarke model)[23,24,43] or (2) a class of high-temperature, disordered, multilayer adsorbates (the Cannon model).[44] Luo *et al.* further observed the metallic counterparts[10,45,46] and free-surface counterparts[47] to these ceramic IGFs, thereby establishing a broader framework to understand these 2-D interfacial phases.

The occurrences of first-order GB phase-like transitions were evident in Si-Au,[48] TiO$_2$-CuO-SiO$_2$,[49] and elemental Cu,[50] amongst other systems.[51] Several examples of 2D interfacial phases are shown in Figure 2.

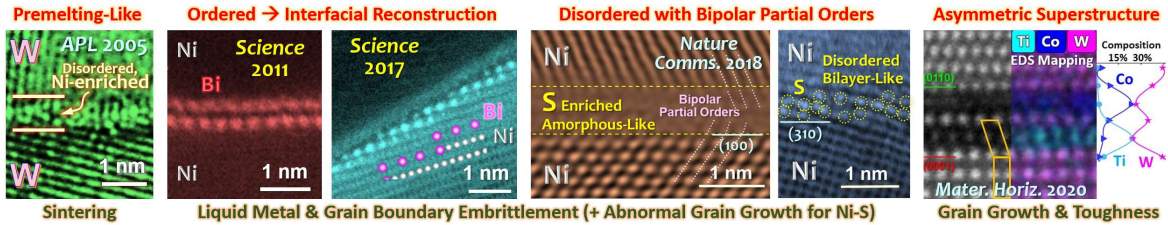

**Figure 2.** Representative 2-D interfacial phases (also called 'complexions'[14,28,52,53]), which can affect or control a broad range of materials properties. (Images are adapted from Refs. [10,54-57].)

In 2006, Tang, Carter, and Cannon[25,26] introduced the term 'complexions' to represent such thermodynamically 2-D interfacial phases; see terminology discussion in §2.3. In 2017, a series of discrete GB complexions were observed in doped Al$_2$O$_3$ by Dillon and Harmer[58-61] and subsequently in other materials.[7,46,48,62-65] These Dillon-Harmer complexions can be considered as derivatives of IGFs with discrete thicknesses of 0, 1, 2, 3, $x$, and $+\infty$ atomic layers.[40,55,62,66] Other complex GB complexions with interfacial reconstructions (*i.e.*, superstructures; see, *e.g.*, an asymmetrical interfacial superstructure at the Ti- and Co co-segregated WC GB in the most right panel in Figure 2 [57]) have also been observed.[54,57,67]

As illustrated in Figure 3, the formation and transition of 2-D interfacial phases (*a.k.a.* complexions) at GBs are of broad importance to materials science. The discovery of interfacial phase-like behaviors had provided new insights towards the understandings of a spectrum of long-standing scientific mysteries, *e.g.*, origins and atomic mechanisms of activated sintering of ceramics and refractory metals,[6-12] liquid metal embrittlement of Ni-Bi and Al-Ga [54,55,68-70] as well as the classical GB embrittlement of Bi *vs.* S-doped Ni (Figure 2),[54-56] and abnormal grain growth in Al$_2$O$_3$ and Ni-S.[56,58,59] IGFs and other GB complexions are also known to affect the toughness, strength, fatigue, and wear resistance of Si$_3$N$_4$, SiC, and Al$_2$O$_3$ and other ceramics,9,24,25,28,46,179,180 the hot strength and creep and oxidation resistance of various structural ceramics,[71-79] superplasticity of zirconia,[80] grain growth and mechanical properties of WC-based cermets,[57,81-83] the stability and mechanical properties of nanocrystalline alloys,[84-94] corrosion of synroc,[95] the electrical resistance of ruthenate thick-film resistors,[96] the coercivity of Nd-Fe-B magnets,[97] the non-linear I-V character of ZnO-based varistors,1,43,44 the critical current of YBCO superconductors,[98] the ionic conductivity of solid electrolytes,[99-101] performance of various battery electrode materials,[102-105] amongst other structural and functional properties.[13,14,28,53,106]

We propose to develop the GB counterparts to the bulk phase diagrams. Since phase diagrams are an essential tool for materials scientists, the capability of computing their GB counterparts can lead to broad scientific and technological gains. This perspective article discusses the models and methods to compute such GB diagrams.

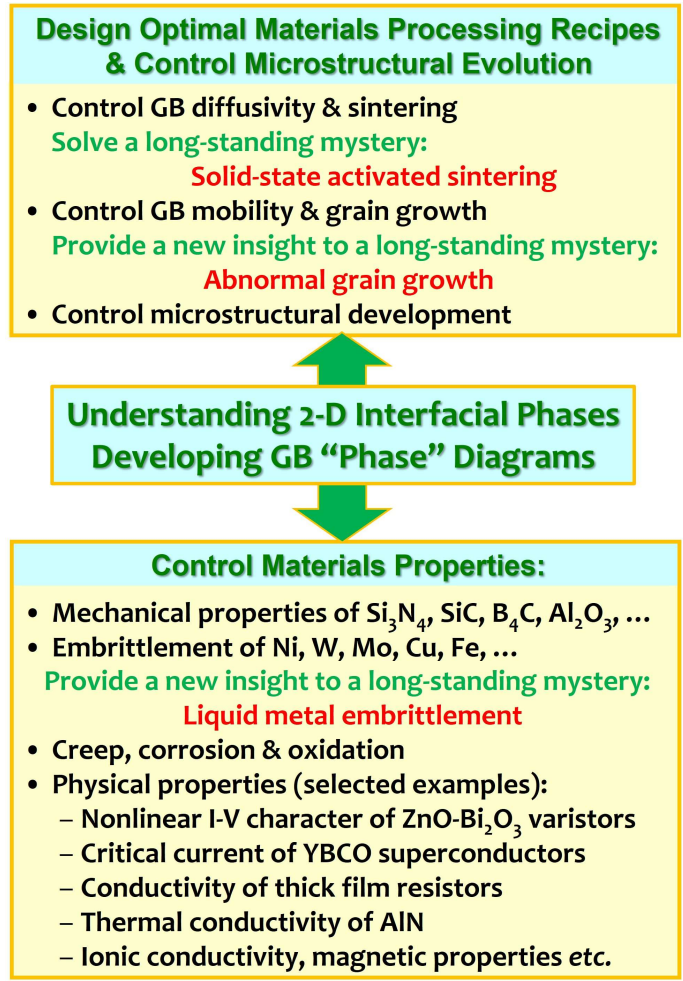

**Figure 3.** Broad scientific and technological impacts of understanding 2-D interfacial phases (complexions) and developing the GB counterparts to bulk phase diagrams.

## 2. A Transformative Scientific Goal: Computing GB Diagrams

### 2.1 Basic Concepts of Interfacial Transitions

To understand the physical origins of GB transitions, we can start from discussing the chemical and structural aspects separately. While they are often coupled in multicomponent materials, simplified models can be built to understand the basic concepts, as well as predict useful trends. Here, let us first examine GB adsorption (*a.k.a.* "segregation"; noting that these two terms are equivalent in thermodynamics and used interchangeable) transitions. The simplest GB adsorption (segregation) model is represented by the Langmuir-McLean isotherm:[107,108]

$$\frac{\Gamma}{\Gamma_0 - \Gamma} = \frac{X_{bulk}}{1 - X_{bulk}} \cdot e^{\frac{-\Delta g_{ads}}{kT}} \qquad (1)$$

where $\Delta g_{ads}$ is the Gibbs free energy of adsorption (defined as a negative value for positive adsorption), $X_{bulk}$ is the bulk fraction of solute, $\Gamma_0$ is the number of adsorption sites at the GB, $\Gamma$ is GB adsorption amount, $k$ is Boltzmann constant, and $T$ is absolute temperature. The Langmuir-McLean model assumes that $\Delta g_{ads}$ is a constant (*i.e.*, no adsorbate-adsorbate interaction, as both the GB and the bulk phase are treated as ideal solutions). Consequently, there is no GB adsorption transition in the Langmuir-McLean isotherm (Figure 4(A)). Subsequently, the Fowler-Guggenheim model[109] (initially proposed for surface adsorption) introduced a parameter $\alpha_{Fowler}$ to represent adsorbate-adsorbate interaction in $\Delta g_{ads}$:

$$\Delta g_{ads} = \Delta g_{ads}^{(0)} + z_1 \alpha_{Fowler} \frac{\Gamma}{\Gamma_0} \qquad (2)$$

Figure 4(A) plots GB adsorption *vs.* bulk composition curves for the Fowler-Guggenheim model. Notably, a first-order adsorption transition occurs for strong adsorbate-adsorbate attraction when

$$\frac{-z_1 \alpha_{Fowler}}{4kT} \equiv \frac{\Omega_{GB}}{2RT} > 1 \qquad (3)$$

where $z_1$ is the coordination number at the GB and $R$ is gas constant. Since the GB pair-interaction parameter $\omega_{GB} \approx -\alpha_{Fowler}/2$, this condition is equivalent to that the effective GB regular-solution parameter $\Omega_{GB} = z_1 N_A \omega_{GB} > 2RT$, which suggests a phase separation at the GB (similar to the criterion of bulk phase separation at $\Omega_{bulk} > 2RT$ in the classical regular-solution model). In the materials science field, Hart first proposed such a GB adsorption transition,[15,16] and Hondros and Seah further elaborated it based on the Fowler-Guggenheim model.[17,18]

In 1977, Cahn proposed his famous critical point wetting model for a binary liquid system with a miscibility gap ($\Omega_{bulk} > 0$, which exhibits a bulk phase separation at $T < T_C = \Omega_{bulk}/2R$).[20] As shown in Figure 4(B), this model predicted a prewetting line, representing first-order adsorption transitions (based on a similar physical origin), which terminates at a surface critical point. Perhaps Cahn was also the first researcher who plotted the interfacial transition line and interfacial critical point in a bulk phase diagram, which motivated us to compute similar GB diagrams.

Similar to surface premelting (Figure 1), GB premelting may be a common type of GB structural transition with interfacial disordering. In 1989, Hsieh and Balluffi reported an *in situ* hot-stage transmission electron microscopy (TEM) experiment that concluded that GB premelting likely occurs for pure Al, but only above $0.999 T_{melting}$.[110] In 2005, the occurrence of GB premelting in colloidal crystals was reported (Figure 4(C)).[111] Nonetheless, the characterization of GB premelting in unary materials remains difficult. Interestingly, impurity-based, liquid-like IGFs have been found to form below bulk solidus lines (*e.g.*, Figure 5(A)),[10,45-47,112-115] which can be interpreted as coupled GB premelting (structural disordering) and prewetting (adsorption) using a generalized Cahn model.[20,25] Here, we recognize that 2-D interfacial phases and GB transitions can be more complex. Not only the structural and adsorption transitions are often coupled, but also interfacial reconstructions can occur, which can lead to new interfacial orders or different symmetries.[54,57,67] In addition, GBs have five macroscopic (crystallographic) degrees of freedom (DOFs). Thus, various methods based on thermodynamic models, atomistic simulations, and machine learning are needed to compute GB diagrams with tradeoffs in their accuracy, efficiency, and robustness.

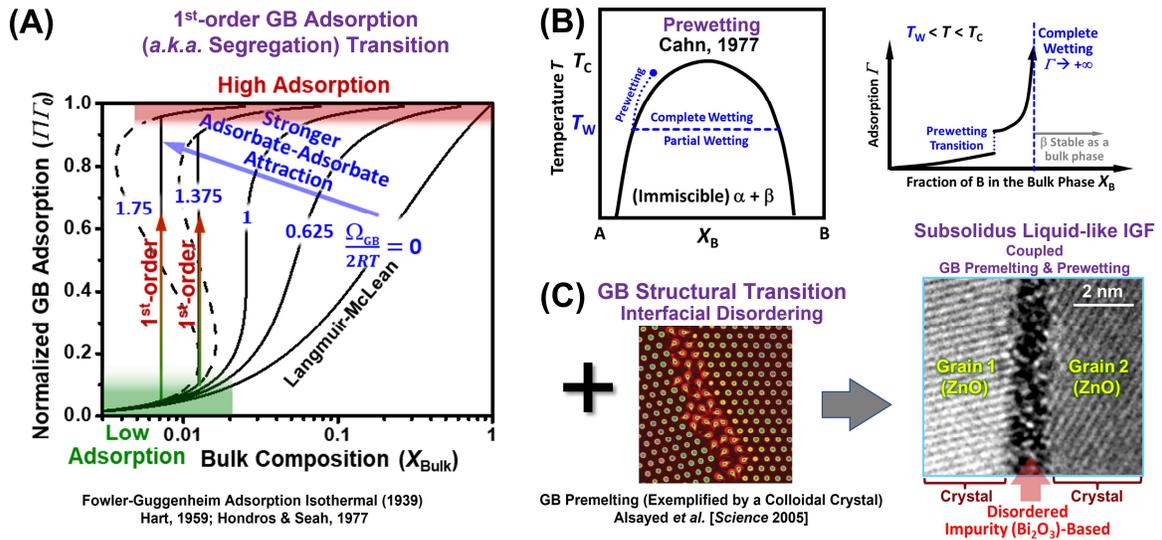

**Figure 4.** Basic concepts underpinning interfacial transitions. **(A)** First-order adsorption transitions occur in Fowler-Guggenheim model for strong adsorbate-adsorbate attraction ($\Omega_{GB}/2RT = -z_1\alpha_{Fowler}/4kT > 1$). **(B)** Cahn's critical point wetting model derived a prewetting adsorption transition that corresponds to a discontinuous jump in adsorption with increasing bulk composition.[20] Cahn also plotted the interfacial (prewetting) transition line and interfacial critical point in a bulk phase diagram that motivated subsequent studies to construct similar GB diagrams. **(C)** The formation of impurity-based, liquid-like IGFs at subsolidus temperatures can be understood from coupling prewetting (adsorption) and premelting (interfacial disordering). (The image of GB premelting of a colloid crystal was adapted from Alsayed *et al.*, *Science* 2005.[111])

**2.2 Overview: Motivation, Selected Examples, and Usefulness**

The development of the GB counterparts to bulk phase diagrams is motivated by the following concepts. On the one hand, bulk phase diagrams and calculation of phase diagrams (CALPHAD) methods are among the most useful tools for materials scientists. On the other hand, most engineered materials are polycrystalline, where GBs can often control a variety of properties (Figure 3). Moreover, a series of studies showed that GBs can be treated as 2-D interfacial phases that can undergo phase-like transitions that alter the mechanical and other physical properties (sometimes abruptly, *e.g.*, causing catastrophic failure). The GB thermodynamic states (*i.e.*, the equilibrium profiles of interfacial structures and compositions) and their transitions can also control the materials fabrication processing (*e.g.*, lowering sintering temperatures via forming liquid-like GBs in activated sintering, as shown in Figure 5(A)) and microstructural evolution (*e.g.*, normal and abnormal grain growth), thereby influencing the properties of resultant materials. Thus, the development of the GB diagrams and their computing methods can enable new ways to tailor the processing and properties of various engineered materials.

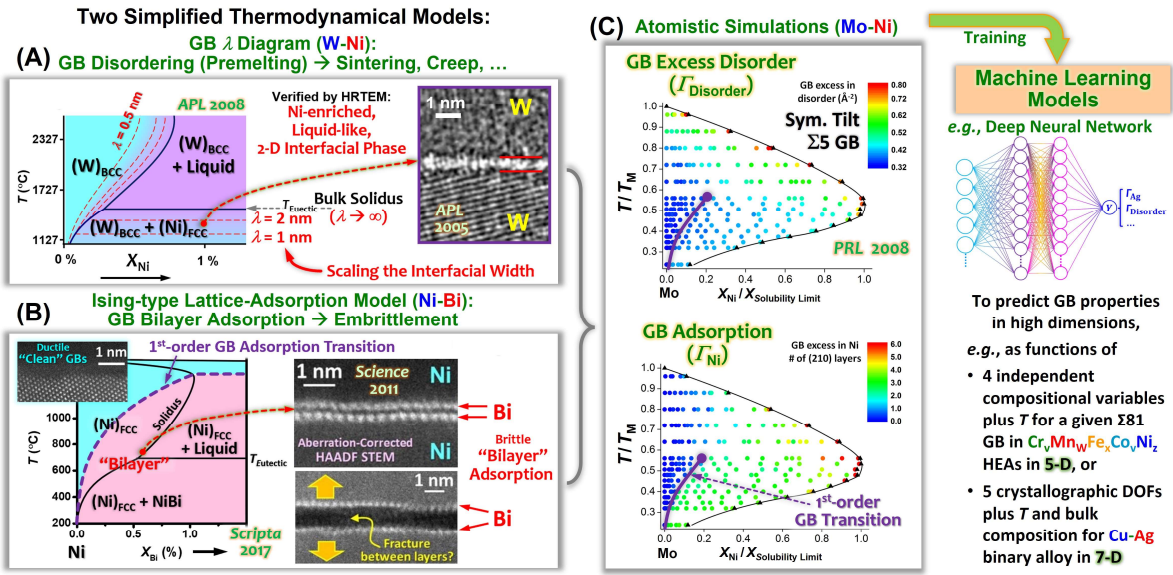

**Figure 5.** Selected examples of computed GB diagrams and the evolution of different models and methods. **(A)** Computed GB $\lambda$ diagram for Ni-doped W to forecast high-temperature GB disordering and related activated sintering behaviors.[10,116] **(B)** Computed GB adsorption diagram for Bi-doped Ni to forecast formation of Bi-based bilayer adsorption and related GB embrittlement behaviors (*i.e.*, the formation of ductile *vs.* brittle GBs).[55,117] **(C)** Computed GB adsorption and GB excess disorder diagrams for a special symmetrically tilt $\Sigma 5$ GB in Ni-doped Mo using more accurate atomistic simulations.[118] Furthermore, atomistic simulations data can be used to train machine learning models to expand the prediction power to forecast GB properties in high dimensional spaces.

We can compute GB diagrams based on thermodynamic models, which can predict useful trends despite the simplification. Let us discuss two selected examples below:

- Figure 5(A) shows a computed GB $\lambda$ diagram for Ni-doped W (based on a continuum phenomenological thermodynamic model) to forecast high-temperature GB disordering (*i.e.*, the formation of liquid-like GBs) and related activated sintering behaviors (and potentially also creep resistance).[116]
- Figure 5(B) shows a GB adsorption diagram for Bi-doped Ni (based on an Ising-type statistical thermodynamic model without considering interfacial structural changes) to forecast GB adsorption (the formation of Bi-based bilayer adsorption in this specific case) and related GB embrittlement behaviors (*i.e.*, the formation of ductile *vs.* brittle GBs as a function of temperature and Bi content).[117]

We will further discuss these two types of simplified thermodynamic models that treat the GB disordering and GB adsorption separately in §3.1 and §3.2, respectively.

Subsequently, methodologies were developed to use hybrid Monte Carlo and molecular dynamics (hybrid MC/MD) atomistic simulations to compute more realistic GB 'phase' diagrams, including, but not limited to, GB adsorption and GB excess disorder diagrams (elaborated in §5). One example is shown in Figure 5(C) for a special symmetric tilt Σ5 GB in Ni-doped Mo.[118] Atomistic simulations are not only more accurate, but also able to predict new interfacial phenomena (*e.g.* the broken symmetry for this symmetric tilt GB, that will discussed in §5).[118] Furthermore, atomistic simulations data can be used to train machine learning models to expand the prediction power to forecast GB properties in high dimensions, *e.g.*, GB properties as functions of (*i*) four independent compositional DOF plus temperature in 5-D for a given GB in high-entropy alloys [119] or (*ii*) five crystallographic DOFs plus and bulk composition in 7-D for Cu-Ag,[120] which will be elaborated in §6.

More examples of computed GB diagrams computed by thermodynamic models, atomistic simulations, and machine learning and their experimental validations will be discussed in detail subsequently. Such GB diagrams can be used for (*i*) optimizing fabrication pathways to utilize desired interfacial structures during processing to control microstructural evolution, (*ii*) designing heat treatment recipes to adjust interfacial structures to improve mechanical or other physical properties, and (*iii*) forecasting GB-controlled, high-temperature material properties. The necessity and usefulness of developing such GB diagrams are demonstrated by studies of solid-state activated sintering in both metallic and ceramic systems.[5,8,45] Since nanoscale, liquid-like interfacial phases can form well below the bulk solidus lines (at subsolidus temperatures) and result in enhanced sintering behaviors similar to liquid-phase sintering,[39] bulk phase diagrams are insufficient for designing activated sintering protocols. Instead, the computed GB $\lambda$ diagrams are proven useful for predicting onset sintering temperatures and trends. See Figure 5(A) for one example and further elaboration in §3.1.

**2.3 Terminology Discussion**

Our general goal is to compute diagrams of various GB properties. The majority of the past work computed diagrams of GB thermodynamic properties such as GB adsorption and GB excess disorder, which can often indicate GB transitions where GB properties change abruptly. In this regard, some of these GB diagrams are indeed the GB counterparts to bulk phase diagrams. However, some other models do not or cannot predict first-order GB transitions. The GB $\lambda$ diagram shown in Figure 5(A) and discussed further in §3.1 only predict trends for high-temperature disorder and related sintering, creep, and other properties. The lattice model discussed in §3.1 can predict first-order GB transitions and critical points, but it ignores structural changes (so the predictions are not always accurate or realistic). Even atomistic simulations sometimes cannot capture GB structural transitions because of their precisions and numerical noises. Thus, some of the GB diagrams shown here are not rigorous GB 'phase' diagrams with well-defined and accurate GB transition lines and critical points. Nonetheless, they are meaningfully and useful as long as they can predict trends that can guide experiments or suggest new interfacial phenomena that can be verified by experiments.

In addition, Tang, Carter, and Cannon introduced the term 'complexions' to represent 2-D interfacial phases based on an argument that they are not rigorously "Gibbs phases".[25,26] An additional goal is to differentiate them from thin layers of secondary bulk phases precipitated at GBs (that are often called "GB phases" in literature). What is the better terminology remains highly controversial, but somewhat subjective (with no significant dispute on the underlying physics).

When the terminology "2-D interfacial phases" are used, we emphasize that they are thermodynamically 2-D, *i.e.*, the compositional and structural profiles along the third dimension are thermodynamically determined (*e.g.*, they have thermodynamically determined "equilibrium" thickness or effective interfacial width that cannot be varied at a thermodynamic equilibrium). Here, "2-D" also emphasizes that they are not precipitated GB phases with arbitrary thickness. In this regard, we recognize that the terminology "complexion" has its advantage.[14] However, both terminologies were/are used previously and currently, and they will likely co-exist in the scientific literature in future. In this perspective article, both terminologies are referred and discussed for the sake of building better connections with all existing literature, and we aim at reducing confusions.

For reasons discussed above, we put 'phase' in quotation marks when the term "grain boundary 'phase' diagrams" or "GB 'phase' diagram" is used (where single quotation marks are used for special terminologies). We can alternatively use "complexion diagrams".

## 3. Two Thermodynamic Models for Predicting GB Disordering *vs.* Adsorption

Here, we will first discuss two thermodynamic models: (***i***) a phenomenological interfacial thermodynamics model to forecast high-temperature interfacial disordering and (***ii***) a lattice statistical interfacial thermodynamics model to predict GB adsorption. Although they are simplified, both models have been proven robustly useful for predicting trends that have been verified by various experiments.

### 3.1 GB $\lambda$ Diagrams

A phenomenological interfacial thermodynamic model were formulated by combining the premelting model[4] and the Clarke model[23,24,43]. Here, a nanoscale disordered complexion (an intergranular film) is treated as a confined liquid-like interfacial film with modified thermodynamic properties. Its excess grand potential as a function of the film thickness ($h$) can be written as:

$$\sigma^{x}(h) = 2\gamma_{cl} + \Delta G_{amorph}^{(vol)} \cdot h + \sigma_{interfacial}(h) , \qquad (4)$$

where $\gamma_{cl}$ is the crystal-liquid interfacial energy, $\Delta G_{amorph}^{(vol)}$ is the free energy per unit volume for forming an undercooling liquid from the equilibrium solid phases, and $\sigma_{interfacial}(h)$ is an interfacial potential ($\sigma_{\text{interfacial}}(+\infty) = 0$; $d\sigma_{\text{interfacial}}/dh$ is the Derjaguin disjoining pressure) that represent the sum of all interfacial interactions.

As shown in Figure 6, a premelting-like interfacial film can form at $T < T_{\text{solidus}}$ if:

$$-\Delta\gamma \cdot f(h) > \Delta G_{\text{amorph}}^{(\text{vol})} h \ , \tag{5}$$

where $-\Delta\gamma$ ($\equiv \gamma_{\text{gb}}^{(0)} - 2\gamma_{\text{cl}} > 0$) is the reduction in the interfacial energy by replacing a high-energy GB ($\gamma_{\text{gb}}^{(0)} \equiv \sigma^x(0)$), which is the "dry" GB energy without temperature-induced disordering) with two low-energy crystal-liquid interfaces ($2\gamma_{\text{cl}}$), and $f(h)$ ($\equiv 1 + \sigma_{\text{interfacial}}(h)/\Delta\gamma$) is a dimensionless interfacial coefficient ($f(0) = 0; f(+\infty) = 1$). This liquid-like interfacial film adopts an equilibrium thickness ($h_{\text{eq.}}$) that corresponds to the minimum of Eq. (4) and satisfies $d\sigma^x(h)/dh = 0$, which can be interpreted as a balance of attractive and repulsive interfacial pressures (akin to the Clarke model).[23,24,115]

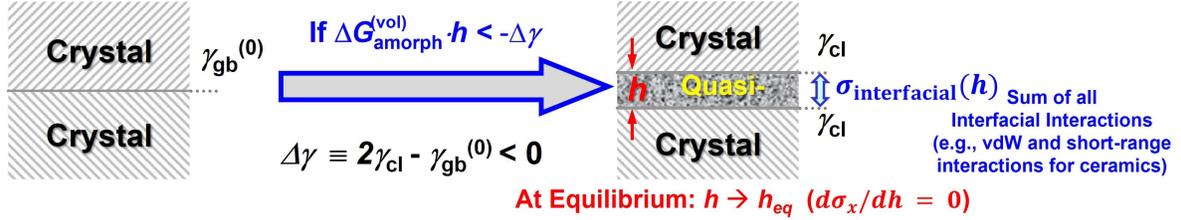

**Figure 6.** A liquid-like interfacial film can be stabilized at the GB below the bulk solidus temperature if the reduction in the interfacial energy upon replacing the "dry" GB with two crystal-liquid interfaces ($\gamma_{\text{gb}}^{(0)} - 2\gamma_{\text{cl}} \equiv -\Delta\gamma$) is greater than the free-energy penalty for forming an undercooled liquid film ($\Delta G_{\text{amorph.}}^{(\text{vol})} \cdot h$).

Based on the above physical principle and Eq. (5), we can define and quantify a thermodynamic parameter, $\lambda$, as:

$$\lambda \equiv \text{Max}\{-\Delta\gamma / \Delta G_{\text{amorph}}^{(\text{vol})}\}, \tag{6}$$

which represents the thermodynamic tendency for average general GBs to disorder. Here, we typically select the film composition that maximizes the $\lambda$ value. Other simplified conventions, *e.g.*, selecting the liquidus composition, lead to similar trends (but slightly lower values) with less computations. Here, $\Delta G_{\text{amorph}}^{(\text{vol})}$ can be quantified by bulk CALPHAD databases, and the interfacial energies can be obtained from statistical thermodynamic models (*e.g.*, Miedema-type models for transition metal alloys[121,122]), experimental values, or density functional theory (DFT) calculations.

Subsequently, we can extend bulk CALPHAD methods to GBs to construct "GB lambda ($\lambda$) diagrams". Figure 5(A) was the first such GB $\lambda$ diagram computed and reported for Ni-doped W, with experimental validation.[116] Additional representative computed GB $\lambda$ diagrams are shown in Figure 7.

These computed GB $\lambda$ diagrams have been validated systematically by experiments firstly in binary refractory alloys (that are classical activated sintering systems):

(1) model predictions were corroborated with direct high-resolution TEM (HRTEM; see, *e.g.*, Figures 5(A), 7(D), and 7(F)) and Auger analysis[7,39,45,46,113],

(2) computed GB $\lambda$ diagrams (with no free parameters) correctly predicted the onset sintering temperatures and trends for W with a series of transition metals as sintering aids (ranked as Pd > Ni > Co ≈ Fe >> Cu),[7,39,123] and

(3) the estimated temperature- and composition-dependent GB diffusivity map for Ni-doped Mo correlated well with the computed GB $\lambda$ diagram (Figure 7(B)).[7]

Furthermore, the successes of computing useful GB $\lambda$ diagrams have been extended from simpler metallic alloys to more complex ceramics. Figure 7(D) shows the computed GB $\lambda$ diagram for CuO-doped $TiO_2$, which was directly verified by HRTEM.[6] It can forecast the trends of CuO-activated sintering of $TiO_2$.[6]

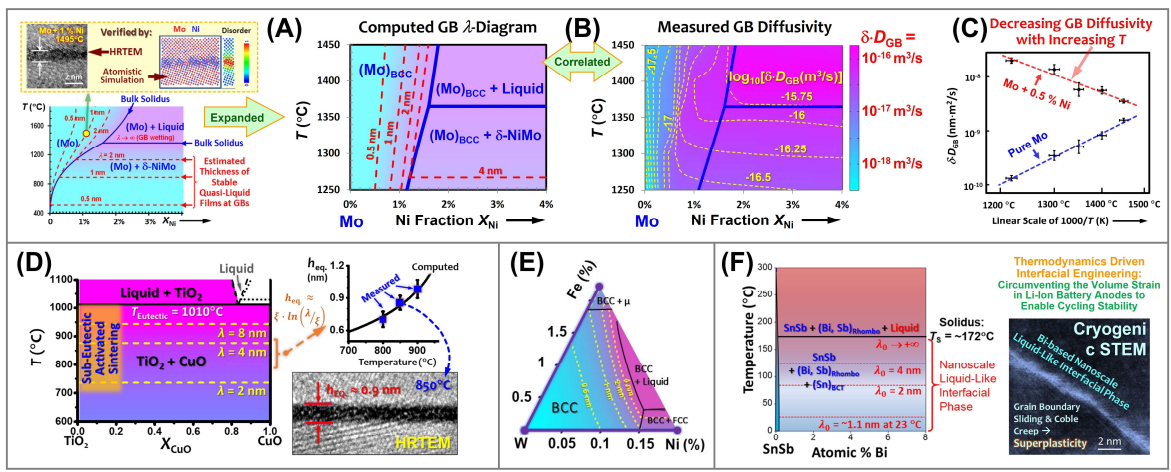

**Figure 7.** Representative computed GB $\lambda$ diagrams for (A-C) Ni-doped Mo,[7] (D) CuO-doped $TiO_2$,[6] (E) Ni and Fe co-doped W,[124,125] and (F) Bi-doped SnSb.[105] **(A)** A binary computed GB $\lambda$ diagram for Ni-doped Mo, with direct validation with high-resolution TEM (HRTEM) and atomistic simulations.[7,46] **(B)** An expanded version of the GB $\lambda$ diagram, which correlate well with GB diffusivity map estimated from sintering experiments.[7] **(C)** Experimental validation of a counterintuitive prediction of reduced GB diffusivity with increasing temperature for 1 at % Ni doped Mo.[126] Representative GB $\lambda$ diagrams for **(D)** a ceramic system[6] and **(E)** a ternary alloy,[124,125] both of which predicted trends in sintering verified by experiments. **(F)** A computed GB $\lambda$ diagram plotted in the isopleth of SnSb-Bi, predicting the formation of liquid-like GBs and room-temperature superplasticity that was successfully used to design Li-ion battery anodes exhibiting significantly improved cycling stability.[105]

This phenomenological interfacial thermodynamic model was also successfully extended to construct GB $\lambda$ diagrams for multicomponent alloys; see, *e.g.*, Figure 7(E) for an isothermal section of Fe and Ni co-doped W[124,125] and Figure 7(F) for an isopleth of SnSb-Bi.[105] Here, a thermodynamic framework and algorithm were developed for computing multicomponent GB $\lambda$ diagrams.[125] Key thermodynamic parameters that control the interfacial segregation and disordering behaviors have been identified and systematically examined.[124] Ternary and quaternary GB $\lambda$ diagrams have been computed and used to forecast the sintering behaviors that were subsequently verified by experiments.[125]

Notably, the model and computed GB $\lambda$ diagrams can also predict counter-intuitive phenomena that were subsequently verified experimentally. For example, an earlier study predicted decreasing GB diffusivity with increasing temperature from 1200 °C to 1500 °C for 1 at.% Ni-doped Mo (that remains single BCC phase in this temperature region), which was subsequently verified by experiments (Figure 7(C)).[126] This counter-intuitive phenomenon stems from the retrograde solubility of Ni in Mo in this temperature region, which results in an increasing penalty to form the metastable liquid with increasing temperature.

In a most recent study, a computed GB $\lambda$ diagram for SnSb-Bi (Figure 7(F)) predicted the formation of liquid-like GBs at room temperature, which may lead to superplasticity.[105] This unusual phenomenon was successfully used to design nanocrystalline Li-ion battery anodes with significantly improved cycling stability (but without nanoporosity in 99% dense micrometer-sized particles).[105]

In brief, GB $\lambda$ diagrams are not yet rigorous GB counterparts to bulk phase diagrams with well-defined transition lines and critical points, but they are robustly useful for forecasting high-temperature GB disordering and related trends in sintering and other phenomena such as superplasticity (including predicted counterintuitive or unusual phenomena), which have been validated by a spectrum of experiments.

**3.2 GB Adsorption Diagrams from a Lattice Model**

In a second approach, an Ising-type statistical thermodynamic model was used to construct GB adsorption diagrams. Here, a useful lattice model for multilayer GB segregation in metals was developed and elaborated by Wynblatt and Chatain.[33] Although this lattice type statistical thermodynamic model only considers GB adsorption (*a.k.a.* segregation) without interfacial disordering or any other GB structural transitions, they can predict GB adsorption transitions and critical phenomena.

In 2008, Wynblatt and Chatain first applied this model to a hypothetic binary regular solution to construct a GB adsorption diagram showing solid-state wetting and prewetting transitions and a GB critical point.[31]

In a 2021 report,[127] a systematics of GB adsorption (*a.k.a.* segregation) transitions and critical phenomena was derived to expand the classical GB segregation theory. This study showed the occurrence of GB layering *vs.* prewetting transitions and how they are related to one another. Moreover, a normalized segregation strength ($\phi_{seg}$) is introduced to represent several factors that control GB segregation, including strain and bond energies, as well as misorientation for small-angle GBs in a mean-field approximation. The key results are illustrated in Figure 8, which suggest two types of behaviors for strong *vs.* weak segregation/adsorption systems:

- In a strong segregation/adsorption system with a large $\phi_{seg}$, first-order layering transitions occur at low temperatures, producing a series of discrete interfacial phases (akin to the Dillon-Harmer complexions,[58-61] but with even numbers of adsorption layers due the symmetry of twist GBs), which become continuous above GB roughing temperatures.

- With reducing $\phi_{seg}$, the layering transitions gradually merge and finally lump into prewetting transitions without quantized layer numbers in weak segregation/adsorption systems, which is analogous to Cahn's critical-point wetting model.

Furthermore, GB adsorption diagrams with universal characters are constructed as the GB counterpart to the classical exemplar of Pelton-Thompson regular-solution binary bulk phase diagrams.[128] This work sets a baseline for understanding the GB adsorption transitions and critical phenomena in binary regular solution systems.[127]

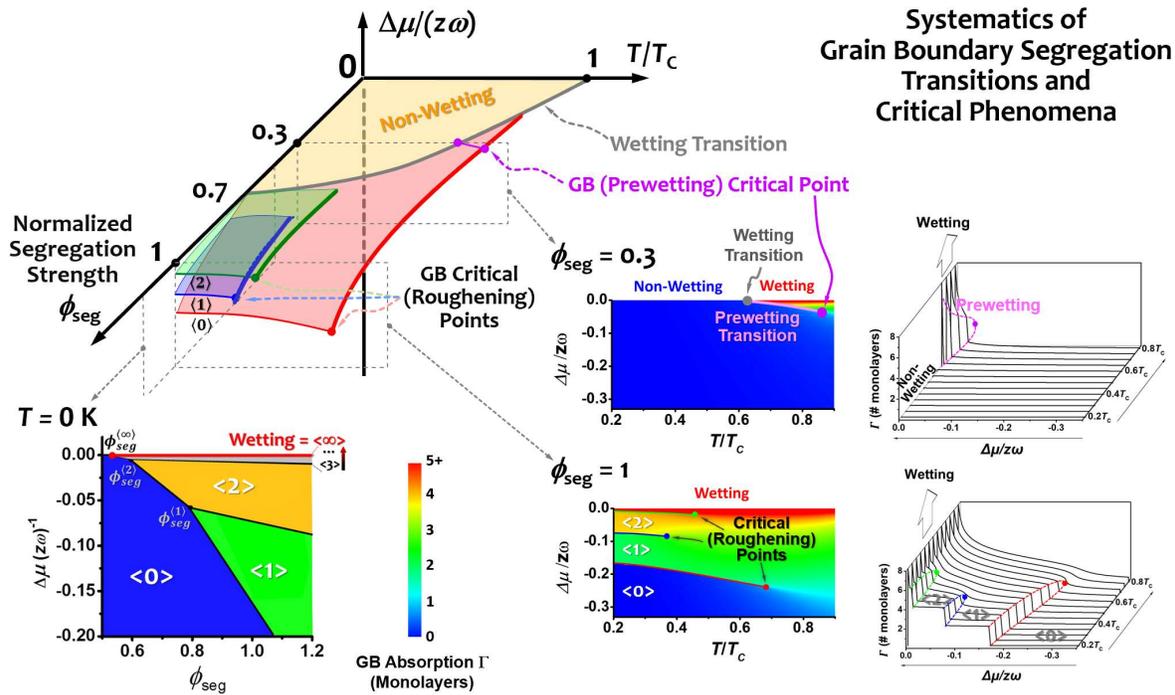

**Figure 8.** A systematics of GB segregation (*a.k.a.* adsorption) transitions and critical phenomena. In a strong segregation system, first-order layering transitions occur at low temperatures and become continuous above GB roughing temperatures. For weaker segregation systems, the layering transitions gradually merge and finally lump into prewetting transitions without quantized layer numbers, akin to Cahn's critical-point wetting model. (Re-plotted after Zhou *et al., Acta Mater.* 2021.[127]).

In addition, the Ising-type lattice-gas model was used to compute GB adsorption (segregation) diagrams for real metallic alloys with results that are consistent with experiments. For example, a GB adsorption diagram was computed for the average general GBs in Bi-doped Ni, as shown in Figure 8(A) in a logarithmic compositional scale to show the full composition region (that is also shown in Figure 4(B) in a linear scale for the Ni-rich portion only).[117] The model was calibrated with DFT calculations. The formation of bilayer adsorption in this system was previously reported and is known to cause GB embrittlement.[55] The predictions of brittle "bilayers" *vs.* ductile "clean" GBs were verified by aberration-corrected (AC) scanning transmission electron microscopy (STEM) high-angle annular dark-field (HAADF) imaging characterization of a series of specimens (Figure 8(A)).[117] The occurrence of first-order transition in the single-phase region was also indicated by prior Auger measurements of fractured GBs.[129]

Figure 9(B) further illustrates the origin of first-order adsorption transitions, as shown in normalized GB energy ($\gamma_{GB}/\gamma_{GB}^{(0)}$) *vs.* bulk Bi composition ($X_{Bi}$) curves and the corresponding computed GB excesses of the solute ($\Gamma$'s). The GB transitions occur when $\gamma_{GB}/\gamma_{GB}^{(0)}$ *vs.* $X_{Bi}$ curves for the "bilayers" and the "clean" GBs intersect. The first-order transitions correspond to the abrupt increases (finite jumps) of absorption, or the associated discontinuities in the slopes in GB energies. The physical origin of this adsorption transition is similar to that shown in Figure 4(A) from the Fowler-Guggenheim model,[17,18] stemming from an adsorbate-adsorbate attraction at the GB.

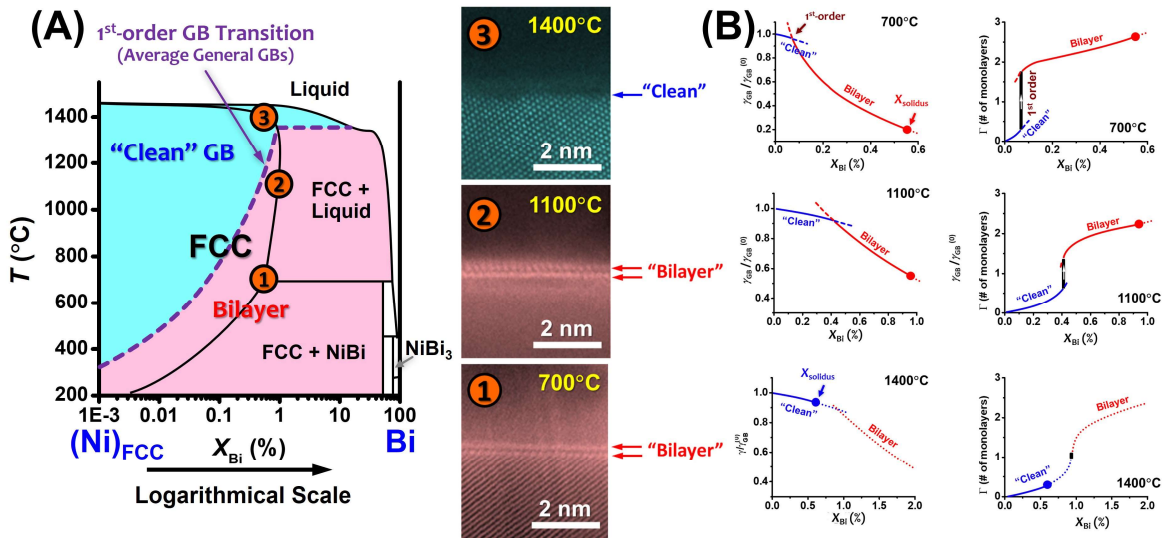

**Figure 9.** **(A)** Computed GB adsorption (segregation) transition diagram for Bi-doped Ni (plotted in full, logarithmic compositional scale), along with AC STEM HAADF images that experimentally verified the predictions for three selected compositions on the solidus line labeled in the diagram. Black solid lines represent bulk phase diagram and purple dashed line indicates the first-order GB transition line for the average general GBs in Bi-doped Ni. **(B)** Calculated normalized GB energy ($\gamma_{GB}/\gamma_{GB}^{(0)}$) *vs.* bulk Bi composition ($X_{Bi}$) curves for Bi-doped Ni and the corresponding computed GB excesses of the solute ($\Gamma$'s) at 700°C, 1100°C, and 1400°C, respectively. The GB transitions occur when $\gamma_{GB}/\gamma_{GB}^{(0)}$ *vs.* $X_{Bi}$ curves for the "bilayers" (red lines) and the "clean" GBs (blue lines) intersect. The first-order transitions correspond to the abrupt increases of absorption or the associated discontinuities in the slopes in GB energies. Dashed lines represent the metastable regions of the complexions. The dotted lines represent the supersaturated (metastable) region beyond the bulk solidus line. It is worth noting that the metastable GB transition is nearly continuous at 1400°C, so that a metastable critical point may exist above 1400°C. (Re-plotted after Zhou *et al., Scripta Mater.* 2017.[117])

## 4. Other Simplified Methods and Models to Construct GB Diagrams and Related Studies

Early studies already sketched GB prewetting and premelting diagrams for Cu-Bi (Bi-doped Cu)[130] and Fe-Si-Zn,[131] which were primarily based experiments of GB chemistry and kinetics measurements for Cu-Bi[130,132-138] and Zn penetration in GBs in Fe-Si alloys[130,131,139-141]. See reviews by Straumal et al.[134] and Rabkin et al.[142] and references therein for earlier studies in this area. The atomistic level interfacial structures had not been directly characterized in those studies. A more recent AC STEM HAADF study showed that general GBs in Cu-Bi form bilayers,[143] akin to those observed in Ni-Bi (Figure 9).

Extending the Cahn critical point wetting model (Figure 4(B)),[20] Tang, Cannon, and Carter used a diffuse-interface (phase-field) model that considered graded crystallinity and orientation profiles (and composition profile in a binary alloy) to compute a GB premelting diagram (as function of temperature and GB misorientation)[26] and a coupled GB premelting and prewetting diagram for a hypothetic binary regular solution[25]. Mishin et al. further elaborated GB premelting like transitions in Ag-doped Cu using a multi-phase-field model that considers composition and crystallinity.[38] Recently, Kamachali and co-workers proposed a density-based phase-field model to compute GB diagrams for Fe-based ternary alloys[144] and other systems.[145,146] This density-based model was parameterized and applied to several alloy systems with some interesting predictions.[144-146] It is noted that crystallinity and density are correlated (not independent) order parameters. It is uncertainty whether density is a better order parameter for the phase-field model (vs. crystallinity used previously by Tang et al. [25,26] and Mishin et al.[38]). In general, future experiments should be conducted to critically examine predictions from these phase-field models.

Luo also developed a thermodynamic model by combining diffuse-interface and lattice-gas models to investigate the interplay of premelting, prewetting, and multilayer adsorption.[40] This model produced GB diagrams showing first-order and continuous coupled prewetting and premelting transitions, critical points, multilayer adsorption, layering and roughening, and complete wetting and drying. It explained the origin of Dillon-Harmer complexions[58,59] in a (perhaps overly) simplified approach. It also showed that the presence of dispersion and electrostatic forces in ceramic materials can appreciably change the GB transitions.

It is worth noting that Wahnström and co-workers have developed an elegant and sphosipcated approach to compute interfacial diagrams based on first-principles methods (considering entropic effects) for coherent transition metal (e.g., Ti, Co, or Cr) doped WC-Co interfaces at finite temperatures.[147-150] This method has not yet been applied to GBs. We also note that extensive studies used DFT based methods to construct the stability diagrams for surfaces and coherent interfaces at 0K (see, e.g., Refs. [151-153] and many others), which are beyond the scope of this perspective article.

## 5. Atomistic Simulations

As discussed in §3, two types of GB diagrams can be computed from simplified thermodynamic models to forecast interfacial (structural) disordering and (chemical) adsorption, respectively. Furthermore, methods to compute more accurate GB 'phase' diagrams were developed by using atomistic simulations, which represent a major advancement in computing GB 'phase' diagrams more rigorously and accurately, which consider both GB adsorption and structural changes (Figure 5).

In the first example, a semi-grand-canonical-ensemble simulation methodology that combines an improved genetic algorithm (GA) with hybrid MC/MD atomistic simulations was developed to construct GB adsorption and excess disorder diagrams for the Ni-doped Mo (Figure 5(C)).[118] Specifically, it combined a modified GA to search for the lowest-energy GB structure through the energy landscape at 0 K with hybrid MC/MD simulations at finite temperatures to predict the equilibrium GB structure as a function of equilibrium temperature ($T$) and chemical potential difference ($\Delta\mu$). Computed GB excess *vs.* bulk composition curves at different temperatures for this GB are shown in Figure 10(A). At low temperatures (*e.g.*, $0.4T_m$), the first-order transitions between "clean" and "bilayer" complexions are observed. The first-order GB transition line ends at a GB critical point, where the GB transition becomes continuous.[118] The computed GB adsorption diagram is shown in Figure 5(C).

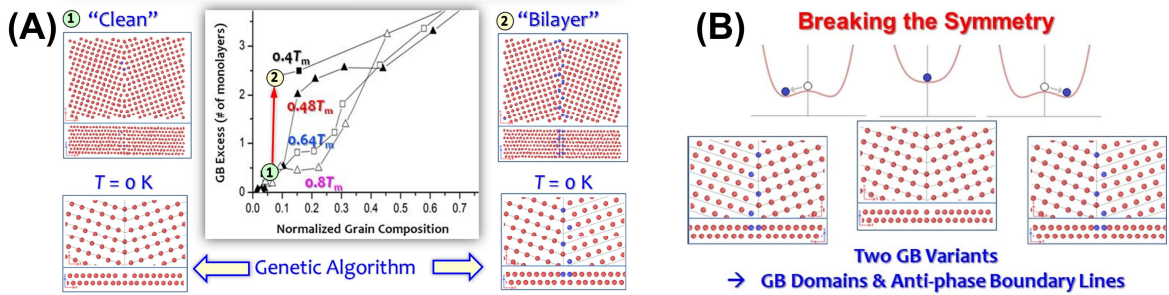

**Figure 10.** The first-order GB transition and critical point predicted from atomistic simulations. The corresponding computed GB adsorption and GB excess disorder diagrams are shown in Figure 5(C). **(A)** The GB excess in Ni *vs.* the normalized bulk composition curved at four different temperatures computed from hybrid MC/MD simulations. The abrupt adsorption transition (at $T = 0.4T_m$) occurs concurrently with a GB structural transition that with a relative translation of the two abutting grains by a distance of $0.3 \cdot a_{bcc}$ along the vertical direction and asymmetric adsorption of the second Ni layer on only one abutting grain. This is clearly illustrated in the atomic configuration simulated by a genetic algorithm at $T = 0$ K. The GB transition becomes continuous at high temperatures, suggesting the existence of a GB critical point. **(B)** The atomistic simulations predicted broken of the mirror symmetry of the symmetric tilt GB, which leads to the co-existence of two variants after the first-order GB transition. (Re-plotted after Yang *et al., Phys. Rev. Lett.* 2018.[118])

As a notable new discovery of interfacial science, it was found that the first-order GB transition can break the mirror symmetry of the Mo Σ5 (210) symmetric tilt GB, producing two variants of asymmetric bilayers (Figure 10(B)).[118] The transition from "clean" to "bilayer" obtained by hybrid MC/MD simulations was reproduced by a modified GA at $T = 0$K without thermal noises, where this symmetry breaking is clearly evident (Figure 10(B)). First-principles DFT calculations was also conducted to confirm the stability of asymmetric bilayers.[118]

In the second example, hybrid MC/MD simulations were used to compute a GB adsorption diagram for a Σ43 (111) twist GB (representing a general twist GB) in Au-doped Si via hybrid MC/MD simulations.[154] The predictions were further verified by DFT calculations. The computed GB adsorption diagram is shown in Figure 11(A).[154] Specifically, hybrid MC/MD simulations have again revealed the occurrence of first-order adsorption transition from nominally "clean" GBs to bilayer adsorption at the Si twist GBs at low temperatures, consistent with a prior experimental observation (as shown in the HAADF image in Figure 11(A)).[48] This first-order GB transition also becomes continuous at high temperatures above a GB critical point (Figure 11(A)). Moreover, hexagonal patterns of Au segregation were identified by hybrid MC/MD simulations as a new prediction, which were further confirmed by DFT calculations (Figure 11(B)).[154]

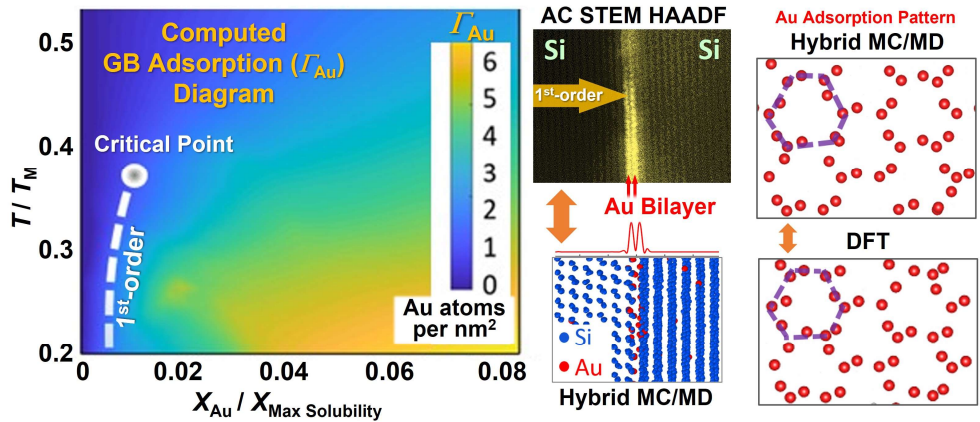

**Figure 11. (A)** GB adsorption diagrams for Au-doped Si Σ43 (111) twist GB computed from hybrid MC/MD atomistic simulations,[154] which suggested a first-order GB transition consistent with a prior bi-crystal experiment.[48] **(B)** STEM HAADF image and the corresponding hybrid MC/MD simulated bilayer configuration. **(C)** The hybrid MC/MD atomistic simulations further revealed hexagonal segregation patterns of Au atoms (at the Si Σ43 and Σ21 (111) twist GBs), which was further confirmed for the Σ21 GB by DFT simulations (as the unit cell for the Σ43 GB is too large for DFT calculations).[154] (Re-plotted after Hu & Luo, *Scripta Mater.* 2019.[154])

## 6. Machine Learning

### 6.2 GB Diagrams for High-Entropy Alloys

High-entropy alloys (HEAs) have attracted great research interests recently.[155-158] However, a fundamental understanding of GBs in HEAs is lacking because of the complex coupling of the segregations of multiple elements and interfacial disordering. By combining large-scale atomistic simulations and machine learning models, a recent study demonstrated the feasibility of predicting the GB properties as functions of four independent compositional degrees of freedoms and temperature in a 5-D space, thereby enabling the construction of GB diagrams for quinary HEAs (Figure 12(A)).[119]

In this study, artificial neural network (ANN), support vector machine (SVM), regression tree, and rational quadratic Gaussian models were trained and tested. The ANN model yielded the best machine learning based predictions.[119] Excellent performance of the ANN model and selected GB diagrams predicted by the ANN model areshown in Figure 12(B).[119]

A data-driven discovery further revealed new coupled segregation and disordering effects in HEAs.[119] For instance, interfacial disordering can enhance the co-segregation of Cr and Mn at CrMnFeCoNi GBs. A physics-informed data-driven model was developed to provide more physical insights and enable better transferability.[119] DFT calculations were used to validate the prediction generality and reveal underlying segregation mechanisms.[119] This study not only provided a new paradigm enabling the prediction of GB properties in a 5-D space, but also uncovered new GB segregation phenomena in HEAs beyond the classical GB segregation models.

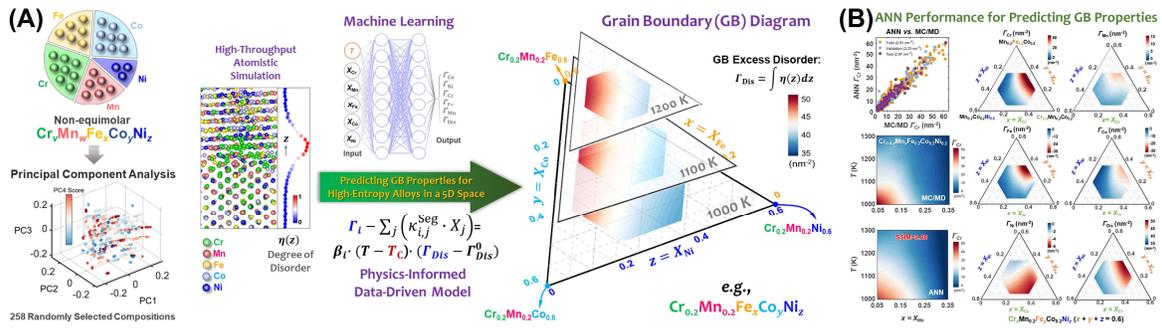

**Figure 12. (A)** Machine learning prediction of GB for non-equimolar quinary $Cr_xMn_yFe_zCo_lNi_m$ quinary alloys. A principal component analysis (PCA) was used to verify the randomness in the selection of 258 HEAs. The equilibrium structure of an asymmetric Σ81 GB (to represent a general GB) in $Co_{0.2}Ni_{0.2}Cr_{0.2}Fe_{0.35}Mn_{0.05}$ at 1000 K obtained by hybrid MC/MD simulations. In total, 1032 such individual hybrid MC/MD simulations were performed for 258 HEAs at four different temperatures to calculate GB excesses of solutes (*i.e.*, $\Gamma_{Cr}, \Gamma_{Mn}, \Gamma_{Fe}, \Gamma_{Co}, \Gamma_{Ni}$) and disorder ($\Gamma_{Dis}$). An artificial neural network (ANN) model was developed for predicting six GB properties. Three other machine learning models have also been trained and tested, while the ANN model is selected because of its best performance. An example of GB disorder diagrams predicted by the ANN model for a ternary $Cr_xMn_yFe_zCo_{0.2}Ni_{0.2}$ ($x + y + z = 0.6$) subsystem is shown. **(B)** ANN performance for predicting GB properties. Parity plot of ANN predictions *vs.* MC/MD simulations for the GB excess of Cr adsorption ($\Gamma_{Cr}$). MC/MD-simulated *vs.* ANN-predicted isopleths of $\Gamma_{Cr}$ diagrams as functions of temperature and Mn bulk composition ($x = X_{Mn}$) for the $Cr_{0.4-x}Mn_xFe_{0.2}Co_{0.2}Ni_{0.2}$ system. Representative ternary isothermal sections of ANN-predicted GB diagrams of $\Gamma_{Cr}, \Gamma_{Mn}, \Gamma_{Fe}, \Gamma_{Co}, \Gamma_{Ni}$, and $\Gamma_{Dis}$ for $Cr_xMn_{0.2}Fe_yCo_{0.2}Ni_z$ ($x + y + z = 0.6$; $x = X_{Cr}, y = X_{Fe}, z = X_{Ni}$) at 1000 K. (Replotted after Hu & Luo, *Materials Horizons* 2022.[119])

## 6.2 Predicting GB Properties in 7-D for a Binary Alloy Considering Five GB DOFs

As we have discussed, constructing GB diagrams as functions of temperature and bulk composition as a general materials science tool on par with phase diagrams representing a potentially transformative research direction. However, a GB has five macroscopic (crystallographic) degrees of freedom or DOFs. It is essentially a "mission impossible" to construct GB diagrams as a function of five DOFs by either experiments or modeling. Here, a recent study combined hybrid MC/MD simulations with a genetic algorithm-guided deep neural network (DNN) model to tackle this grand challenge (Figure 13).[120]

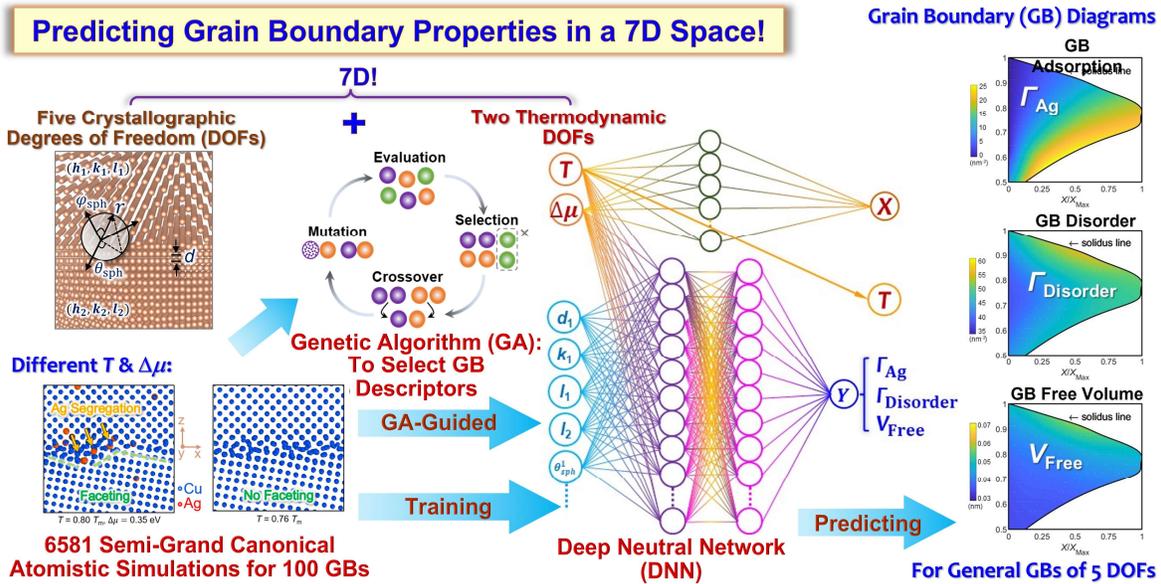

**Figure 13.** Machine learning prediction of bulk composition- and temperature-dependent GB diagrams as a function of five macroscopic degrees of freedom (5 DOFs) for Ag-doped Cu. 6,581 individual constant-$N\Delta\mu PT$ ensemble atomistic simulations were performed for 100 representative GBs to calculate three types of GB diagrams of adsorption ($\Gamma_{Ag}$), excess disorder ($\Gamma_{Disorder}$), and free volume ($V_{Free}$). A genetic algorithm was used selection of significant GB descriptors. The selected significant GB descriptors were used as the input parameters to train, evaluate, and test deep neural network (DNN) models. Schematic diagram of a two-layer single-task DNN with a 15-18-10-1 architecture for predicting GB properties combined with a simplified single-layer ANN for predicting the bulk atomic fraction of Ag is shown. This DNN-based machine learning model enables the forecast of the property diagrams of millions of distinctly different GBs as a function of five macroscopic DOFs, which is otherwise a "mission impossible" to construct by either experiments or modeling. (Replotted after Hu et al., *Materials Toady* 2020.[120])

First, this study performed 6,581 individual isobaric semi-grand-canonical (constant-$N\Delta\mu PT$) ensemble atomistic simulations for 100 representative GBs to calculate GB diagrams of adsorption, excess disorder, and free volume.[120] ~50-100 atomistic simulations are generally required to interpolate one set of three GB diagrams, which takes around 14,000-28,000 core hours of the simulation time per GB. Subsequently, a genetic algorithm was used to select significant GB descriptors. This genetic algorithm is able to re-discover significant parameters that are known to control the properties in each of four classes of GBs (Figure 14(A)). Then, the selected significant GB descriptors were used to as the input parameters to train, evaluate, and test DNN models. Subsequently, this study developed a two-layer single-task DNN for predicting GB properties combined with a single-layer ANN for predicting the bulk atomic fraction of Ag. The input parameters for the all-included DNN model are the significant GB descriptors selected by the genetic algorithm plus two thermodynamic DOFs.

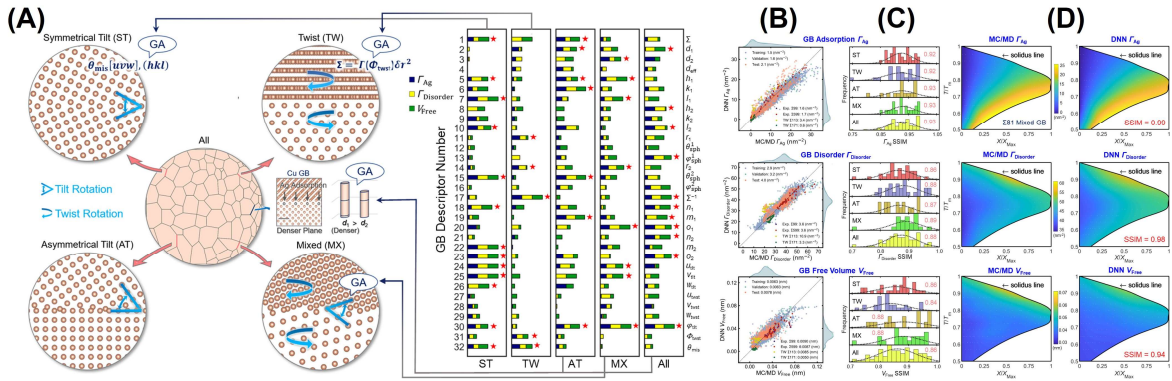

**Figure 14.** Selected results of machine learning prediction of the GB diagrams for general GBs of five DOFs for Ag-doped Cu. **(A)** Genetic algorithm (GA) based variable selection for GB descriptors. The GA scores are plotted for 32 GB descriptors for classes of symmetric-tilt (ST), twist (TW), asymmetric-tilt (AT), and mixed tilt-twist (MX) GBs, and 38 GB descriptors for MX and all four type GBs together (denoted as "All"). The red pentagram stars are used to label GA selected significant descriptors. The most significant GB descriptors selected by the GA include the parameters in the common notation $\theta_{\mathrm{mis}}[uvw](hkl)$ for ST GBs and the those in the characteristic relation $\Sigma = \gamma(\Phi_{\mathrm{twst}})\delta r^2$ for twist GBs. Moreover, the GA finds $d_1$ (of the lower-index plane) and $d_2$ to be the most and second most significant descriptors for AT GBs, as well as all-included GBs, which suggest that GB properties are dominated by the (denser) lower-index plane. **(B)** Performance of deep neural network (DNN) models, shown by parity plots of DNN-predicted values of $\Gamma_{\mathrm{Ag}}$, $\Gamma_{\mathrm{Disorder}}$, and $V_{\mathrm{Free}}$ using the all-included DNN model with an optimized network architecture 15-18-10-1 vs. hybrid MC/MD-simulated values (via atomistic simulations). **(C)** Histogram plots with distribution line (black dotted line) of structural similarity index (SSIM) for characterizing the similarities between MC/MD-simulated GB diagrams and DNN-predicted GB diagrams of $\Gamma_{\mathrm{Ag}}$, $\Gamma_{\mathrm{Disorder}}$, and $V_{\mathrm{Free}}$. Mean SSIMs for each GB type are labeled. **(D)** Comparison of the MC/MD-simulated vs. DNN-predicted GB $\Gamma_{\mathrm{Ag}}$, $\Gamma_{\mathrm{Disorder}}$, and $V_{\mathrm{Free}}$ diagrams using all-included DNN model for a Σ81 mixed GB with boundary planes $(1\bar{1}0)//(7\bar{8}7)$. The SSIMs for characterizing the similarities between MC/MD-simulated and DNN-predicted GB diagrams are labeled. (Replotted after Hu et al., *Materials Toady* 2020.[120])

The DNN prediction is approximately eight orders of magnitude (~$10^8$) faster than atomistic simulations, thereby enabling the construction of the property diagrams for millions of distinctly different GBs of five DOFs.[120] The good performance of the DNN models is shown in Figure 14(B-D). Excellent prediction accuracies have been achieved for not only symmetric-tilt and twist GBs, but also asymmetric-tilt and mixed tilt-twist GBs. The latter are more complex and much less understood, but they are ubiquitous and often limit the performance properties of real polycrystals as the weak links.

In brief, this deep learning model enables the forecast of the GB diagrams of millions of distinctly different GBs as a function of five macroscopic DOFs, which is otherwise impossible to construct by either experiments or modeling. The data-driven prediction of GB properties as function of temperature, bulk composition, and five crystallographic DOFs (i.e., in a 7-D space) opens a new paradigm. We may further extend this methodology to other binary and multicomponent alloys to predict GB properties in high (7+) dimensional spaces.

## 7. From Thermodynamic to Mechanical Properties

Using a classical embrittlement model system Ga-doped Al, a recent study further demonstrated the feasibility of computing temperature- and composition-dependent GB diagrams to represent not only equilibrium thermodynamic and structural characters (Figure 15(D-G)), but also mechanical properties (Figure 15(H-I)).[69]

Hybrid MC/MD simulations were first used to obtain the equilibrium GB structure as a function of temperature and composition.[69] Simulated GB structures were validated by AC STEM (Figure 15(A)) to ensure the validity and accuracy of the atomistic simulations. Subsequently, the hybrid MC/MD simulated diagrams of adsorption or GB excess of Ga ($\Gamma_{Ga}$) and GB excess of structural disorder ($\Gamma_{Disorder}$) are shown in Figure 15(D) and Figure 15(E), respectively.[69] These GB diagrams were computed for an asymmetric Σ81 GB to represent a general GB.

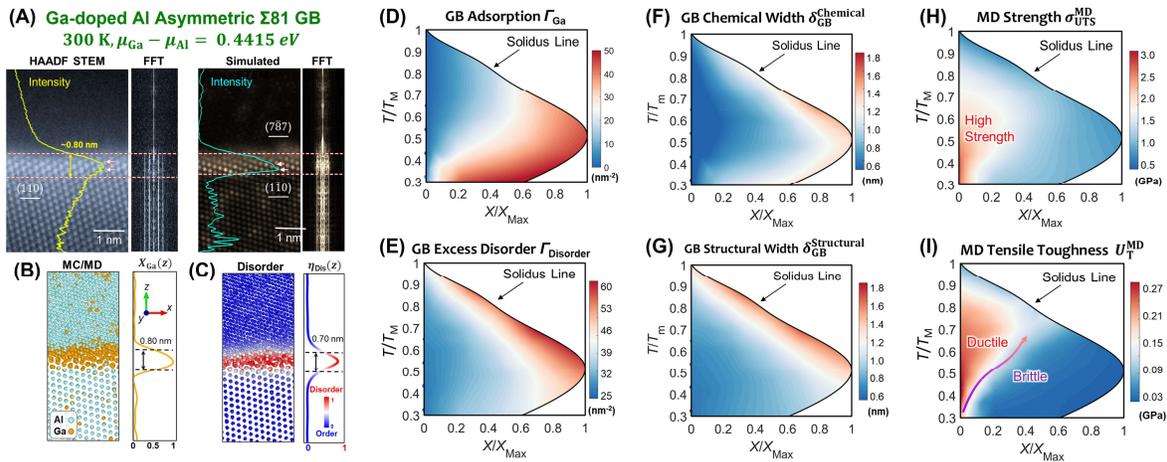

**Figure 15.** Computing GB diagrams of thermodynamic and mechanical properties for Ga-doped Al.[69] **(A)** Validation of the computational approach by comparing an experimental STEM HAADF image with a simulated STEM image of an asymmetric Σ81 GB that bet matches the experiment, based on the equilibrium atomistic structure obtained from hybrid MC/MD simulations. **(B)** The hybrid MC/MD simulated GB structure and the 2D averaged Ga atomic fraction ($X_{Ga}(z); X_{Ga}(\pm\infty) = X$) profile across the GB as function of the spatial variable *z* perpendicular to the GB. **(C)** The corresponding color maps based on computed disorder parameter ($\eta_{Dis}$) on each atom and the corresponding 2D averaged disorder $\eta_{Dis}(z)$ profile across the GB. The hybrid MC/MD simulated diagrams of **(D)** adsorption or GB excess of Ga ($\Gamma_{Ga}$) and **(E)** GB excess of structural disorder ($\Gamma_{Disorder}$). The corresponding computed GB diagrams of the **(F)** effective GB chemical width ($\delta_{GB}^{Chemical}$) and **(G)** effective GB structural (disorder) width ($\delta_{GB}^{Structural}$). Computed GB diagrams of **(H)** MD ultimate tensile strength ($\sigma_{UTS}^{MD}$) and **(I)** MD tensile toughness ($U_T^{MD}$) from the molecular dynamics (MD) tensile tests based on the hybrid MC/MD simulated equilibrium GB structures. (Replotted after Hu & Luo, *npj Computational Mater.* 2021.[69])

Moreover, the interfacial chemical and structural widths were calculated from atomistic simulations, as shown in Figure 15(B) and Figure 15(C), respectively.[69] The corresponding computed GB diagrams of the effective GB chemical width ($\delta_{GB}^{Chemical}$) and effective GB structural (disorder) width ($\delta_{GB}^{Structural}$) are shown in Figure 15(F) and Figure 15(G), respectively, which represent additional GB diagrams of interfacial thermodynamic and structural properties. Notably, these two new types of GB diagrams indicated different trends in how the structural and chemical widths depend in temperature *vs.* bulk composition.

Subsequently, molecular dynamics (MD) tensile tests were performed on the simulated equilibrium GB structures. GB diagrams were constructed to represent MD ultimate tensile strength (Figure 15(H)) and MD tensile toughness (Figure 15(I)), respectively.[69] These represent new types of computed GB diagrams of mechanical properties.

This study suggested a new and promising research direction to investigate GB composition-structure-property relationships via computing GB diagrams of not only thermodynamic and structural properties, but also mechanical and potentially other properties.

## 8. Summary and Outlook

Since bulk phase diagrams are a foundation for modern materials science, we believe that computing their GB counterparts can have equally significant impacts. This perspective article surveyed a series of studies to compute GB diagrams via thermodynamic modelling, atomistic simulations, and machine learning that are complementary. First, we can use two simplified interfacial thermodynamic models to construct 1) GB $\lambda$ diagrams to forecast high-temperature GB disordering and trends in related sintering and other properties and 2) GB adsorption diagrams to predict GB adsorption transitions and critical phenomena. These GB diagrams represent the GB (structural) disordering and (chemical) adsorption aspects, respectively. Subsequently, we can use hybrid MC/MD atomistic simulations to compute more rigorous and accurate GB 'phase' (complexion) diagrams. In addition, we can extend the computed GB diagrams of thermodynamic and structural properties to further include mechanical and other physical properties. Moreover, we can combine machine learning models with atomistic simulations to predict GB properties in high dimensional spaces to greatly expand the prediction power (*e.g.*, as functions of four independent compositional variables and temperature in a 5-D space for a given GB in high-entropy alloys or as functions of five macroscopic DOFs plus temperature and composition for a binary alloy in a 7-D space).

Different models and approaches have their own pros and cons. While thermodynamic models are less accurate with major simplifications, we can use them to predict useful trends robustly and reveal the underlying physics clearly, if (and only if) we apply such in simplified models appropriately. Atomistic simulations are more accurately, but they are computationally much more expensive. Moreover, their accuracies are often limited by the availability of good interatomic potentials. Machine learning models can help expanding the prediction power to tackle high dimensional problems, but they need data input, which are currently been fed by brute-force large-scale atomistic simulations that computationally very expensive and also limited by available interatomic potentials. Further developments of diffuse-interface models and DFT-based methods can complement the existing approaches by filling gaps in the computational accuracy-difficulty-robustness tradeoffs (albeit their own limitations), but more method developments and investigations of real materials with critical experimental validations are needed.

We should also discuss a few emerging fields. The first field is represented by modeling GBs in high-entropy and compositionally complex alloys (HEAs and CCAs)[155-159] and their ceramic counterparts[160,161], and subsequently computing their GB diagrams in high dimensional spaces. Here, the first success of computing GB diagrams of HEAs (via combined atomistic simulations and machine learning) is discussed in §6.1. In general, GBs in HEAs and CCAs are more difficult to model because of the large compositional spaces (where machine learning can be helpful); modeling GBs in HEAs and CCAs is also challenging because of less reliable interatomic potentials and available thermodynamic data for these multicomponent materials. It is even more difficult to predict GB properties (and subsequently compute GB diagrams) for the diversifying classes of high-entropy ceramics (HECs)[160,161] and compositionally complex ceramics (CCCs)[160,162-165], which have attracted substantial and exponentially-growing research interests recently. Here, GBs in high-entropy (and compositionally complex) oxides,[160,162,165-168] borides,[169-172] carbides,[173-175] silicides,[176,177] and fluorides[178] with diversifying crystal structures and different bonding characters can possess exotic yet intriguing thermodynamic and other physical properties. Understanding, predicting, and controlling the GBs in HEAs/CCCs and HECs/CCCs are of critical importance to enable us to attain their full technological potential.

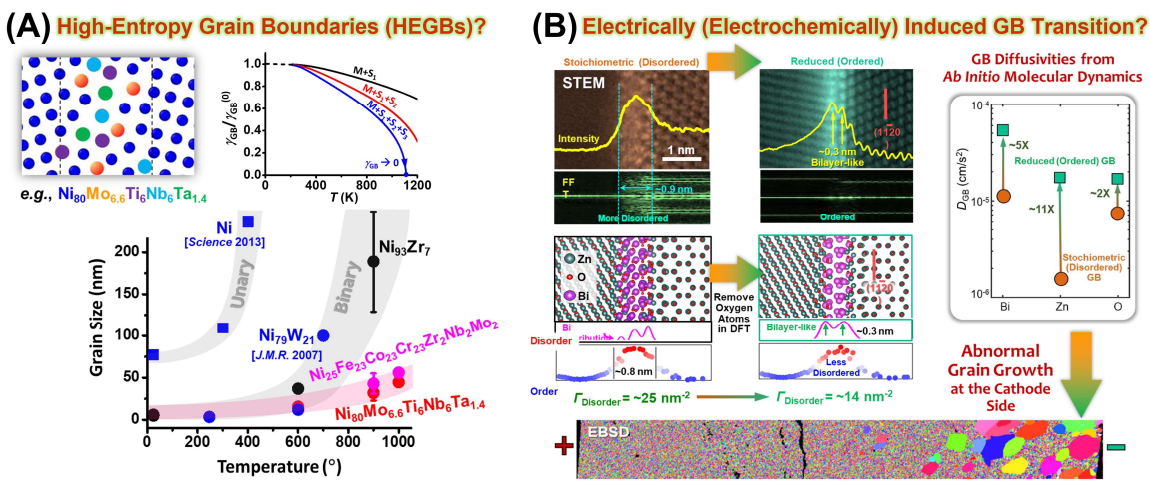

**Figure 16.** Selected emerging areas of interest. **(A)** High-entropy grain boundaries (HEGBs) represent a new typical of 2-D interfacial phases. For example, HEGBs may be utilized to increase the stability of nanocrystalline alloys at high temperatures via both thermodynamic and kinetic effects.[84,124] (Replotted after Zhou & Luo, *Current Opinion* 2016[124] and Zhou *et al.*, *Scripta Mater.* 2016.[84]) **(B)** An applied electric field (or electrostatic and electrochemical potential) represents another knob (dimension) to induce GB phase-like transitions. For example, a recent study demonstrated that an applied electric field can induce a GB disorder-order transition electrochemically, which can enhance GB diffusivity and cause abnormal grain growth.[179] (Replotted after Nie *et al.*, *Nature Communications* 2021.[179])

Here, high-entropy grain boundaries (HEGBs) represent a new type of 2-D interfacial phases. For example, HEGBs can be utilized to increase the stability of nanocrystalline alloys at high temperatures via both thermodynamic and kinetic effects (Figure 16(A)).[84,124] A most recent report also showed the formation of thick amorphous complexion in Cu-Zr-Hf-Nb-Ti and similar quinary nanoalloys.[94] GB phase-like transitions of HEGBs are scientifically interesting. Modeling such HEGBs, including computing GB diagrams to represent their thermodynamic stability and properties, is worth pursuing for future studies.

The second emerging topic is "field-induced" GB transitions (that can in fact be induced by electrostatic or electrochemical potential, instead of the field itself, as suggested by a recent study shown in Figure 16(B)).[179] In thermodynamics, GB transitions (like bulk phase transitions) are often induced by changing a thermodynamic potential, such as temperature, pressure, or chemical potential. However, electrostatic (or electrochemical) potential represents another knob (thermodynamic potential) to induce or control GB transitions. For example, a recent study demonstrated that an applied electric field can induce a GB disorder-order transition in $Bi_2O_3$-doped ZnO electrochemically, which can enhance GB diffusivity and cause abnormal grain growth.[179] We believe this is a general phenomenon with a great potential to open another window or an additional dimension to control GB transitions and properties. How to model such electrically or electrochemically induced GB transitions and how to represent them in computed GB diagrams (by adding electrostatic or electrochemical potential as an additional variable/dimension) represent new scientific problems. Such an effort can lead to exciting new opportunities to tailor GB controlled properties (including, but not limited to, microstructural evolution).

Overall, the field of computing GB counterparts to bulk phase diagrams (and beyond) is still in its infancy stage with an extremely limited number of active researchers, as the computing GB diagrams are highly challenging. Historically, the development of bulk phase diagrams and CALPHAD methods took more than 50 years and efforts of a large number of researchers. Constructing and computing their GB counterparts are perceivably more challenging. However, this is a potentially transformative research direction as GB diagrams can be equally important and useful as bulk phase diagrams.

**Acknowledgement:** The author thanks the National Science Foundation (NSF) Materials Research Science and Engineering Center program through the UC Irvine Center for Complex and Active Materials (Grant No. DMR-2011967) for supporting his current research on fundamental interfacial science of complex concentrated materials. The author also acknowledges partial current supports from the Army Research Office (ARO Grant No. W911NF2210071) for studying HEGBs and the Air Force Office of Scientific Research (AFOSR Grant Nos. FA9550-19-1-0327 and FA9550-22-1-0413) for investigating field-induced GB transitions and microstructural evolution, respectively. The author is grateful for prior supports from a Vannevar Bush Faculty Fellowship as well as the NSF, ONR, AFOSR, and DOE over the past 15 years that made this line of basic research on computing GB diagrams possible. The author thanks many students (particularly Chongze Hu, Naixie Zhou, Shengfeng Yang, Jiuyuan Nie, and Xiaomeng Shi who directly contributed to our work on computing GB diagrams) and collaborators who have helped on various modeling and experimental aspects.